\documentclass[prb,amsmath,amssymb,superscriptaddress,twocolumn]{revtex4-1}
\usepackage{graphicx}
\usepackage{epsfig}
\usepackage{dcolumn}
\usepackage{bm}
\usepackage{rotating}
\usepackage{multirow}
\usepackage[table]{xcolor}
\usepackage[english]{babel}

\newcommand{\be}{\begin{equation}}
\newcommand{\ee}{\end{equation}}
\newcommand{\bea}{\begin{eqnarray}}
\newcommand{\eea}{\end{eqnarray}}
\newcommand{\ba}{\begin{align}}
\newcommand{\ea}{\end{align}}
\newcommand{\nn} {\nonumber}
\renewcommand{\vr} {{\bf r}}

\newcommand{\Tr}{ {\rm Tr} \, }

\def\D{\Delta}

\def\ve{\varepsilon}

\def\l{\lambda}

\def\S{\Sigma}

\def\vf\f{\phi}
\def\vf{\varphi}
\def\F{\Phi}

\def\w{\omega}

\def\bra{\langle}
\def\ket{\rangle}

\def\x{{\rm x}}

\def\Tr{{\rm Tr}\,}

\begin{document}
\widetext 
\title{Static correlation and electron localization in molecular dimers from \\the self-consistent 
RPA and \textit{GW} approximation}
\author{Maria~Hellgren}
\affiliation{International School for Advanced Studies (SISSA), via Bonomea 265, 34136 Trieste, Italy}
\affiliation{Physics and Materials Science Research Unit, University of Luxembourg, 162a avenue de la Fa\"{\i}encerie, L-1511 Luxembourg, Luxembourg}
\author{Fabio~Caruso}
\affiliation{Department of Materials, University of Oxford, Parks Road, Oxford OX1 3PH, United Kingdom}
\author{Daniel~R.~Rohr}
\affiliation{Nano-Bio Spectroscopy Group and European Theoretical Spectroscopy Facility (ETSF), Universidad del País Vasco CFM CSIC-UPV/EHU-MPC DIPC, 20018 San Sebastián, Spain}
\author{Xinguo~Ren}
\affiliation{Key Laboratory of Quantum Information, University of Science and Technology of China, Hefei, 230026, China}
\author{Angel~Rubio}
\affiliation{Max Planck Institute for the Structure and Dynamics of Matter, Hamburg, Germany}
\affiliation{Nano-Bio Spectroscopy Group and European Theoretical Spectroscopy Facility (ETSF), Universidad del País Vasco CFM CSIC-UPV/EHU-MPC DIPC, 20018 San Sebastián, Spain}
\affiliation{Fritz-Haber-Institut der Max-Planck-Gesellschaft, Faradayweg 4-6, D-14195 Berlin, Germany}
\author{Matthias~Scheffler}
\affiliation{Fritz-Haber-Institut der Max-Planck-Gesellschaft, Faradayweg 4-6, D-14195 Berlin, Germany}
\author{Patrick~Rinke}
\affiliation{Fritz-Haber-Institut der Max-Planck-Gesellschaft, Faradayweg 4-6, D-14195 Berlin, Germany}
\affiliation{COMP/Department of Applied Physics, Aalto University, P.O. Box 11100, Aalto FI-00076, Finland}
\date{\today}
\date{\today}
\begin{abstract}
We investigate static correlation and delocalization errors in the self-consistent $GW$ and 
random-phase approximation (RPA) by studying molecular dissociation of the H$_2$ and LiH
molecules. Although both approximations contain topologically identical diagrams, the non-locality and 
frequency dependence of the $GW$ self-energy crucially influence the different energy 
contributions to the total energy as compared to the use of a static local potential in the RPA. 
The latter leads to significantly larger correlation energies which allow for a better 
description of static correlation at intermediate bond distances. The substantial error found in 
$GW$ is further analyzed by comparing spin-restricted and spin-unrestricted calculations.  At large but 
finite nuclear separation their difference gives an estimate of the so-called fractional 
spin error normally determined only in the dissociation limit. Furthermore, a calculation of the dipole 
moment of the LiH molecule at dissociation reveals a large delocalization error 
in $GW$ making the fractional charge error comparable to the RPA.  
The analyses are supplemented by explicit formulae for the $GW$ Green's function and total energy 
of a simplified two-level model providing additional insights into the dissociation limit.
\end{abstract}
\maketitle

\section{Introduction}
Many-body perturbation theory (MBPT)~\cite{fetter}  and Kohn-Sham (KS) density functional theory (DFT) \cite{hohenbergkohn,kohnsham1965,Dreizler} are established as two prominent frameworks in computational electronic structure theory. 
Both are, in principle, exact but in practical calculations a careful choice of approximations has to be made. The central 
quantity to approximate is the exchange-correlation (xc) energy. 
In MBPT, the xc energy is calculated from the interacting one-particle Green's function $G$, which is obtained from 
the non-local and frequency dependent self-energy ($\Sigma$) via Dyson's equation. In contrast, KS-DFT requires a local (i.e. multiplicative) 
and static exchange-correlation potential ($v_{\rm xc}$) and hence approximations to the xc energy must be expressed in 
terms of a non-interacting KS Green's function.  
Only in their simplest form when both $v_{\rm xc}$ and $\Sigma$ are zero, i.e., in the Hartree approximation, the 
MBPT and KS-DFT formalisms are equivalent.

When more advanced approximations to the electron-electron interaction are considered, 
the non-locality and frequency dependence of the self-energy - as opposed to the local 
and static xc potential in KS theory - may lead to qualitative differences in the description 
of the ground state and xc energy in MBPT and DFT. Understanding the origin of these 
differences is essential for advancing the development of new density functional and 
self-energy approximations, and it constitutes the main purpose of this work. 

The  $GW$ approximation\cite{Hedin1965} to MBPT and the random phase approximation 
(RPA) to DFT \cite{Langreth1977} are state-of-the-art approximations for 
first principles excited- and ground-state electronic structure calculations. \cite{Aulbur/Jonsson/Wilkins:2000,Onida/Reining/Rubio:2002,patrick2005,Gorling2010,Eshuis/Bates/Furche:2012,RPAreview} 
Previous works have established their analogies and differences:\cite{hvb07,PhysRevLett.110.146403} 
The xc energy in $GW$ and RPA can be represented in terms 
of topologically identical Feynman diagrams.  Moreover, $GW$ and RPA share a common
total-energy expression,\cite{klein1961,dahlenleeuwen2006,hvb07} i.e., an expression with the 
same functional dependence on the single-particle Green's function.
However, as alluded to in the first paragraph, the RPA energy is 
optimized with respect to Green's functions originating from 
a local KS potential, whereas the \emph{GW} energy is optimized by allowing for free variations of
the Green's function leading to a non-local and frequency dependent self-energy. 
The differences between RPA and $GW$ can thus be ascribed to the framework in which 
they are evaluated.

The performance of perturbative RPA (i.e. the RPA evaluated 
at a non-self-consistent KS Green's function) 
has been studied in several works.\cite{hybertsen85,olsenrpa,kresserpa,furcherpa,ferdirpa,gorlingrpa}
Conversely, the $GW$ approximation, which has mostly been 
used for quasiparticle calculations, has only recently
been explored for the calculation of ground-state properties.\cite{stan,phd/thesis/caruso,PhysRevB.77.115333}
The development of the fully self-consistent 
RPA (scRPA) and $GW$ (sc$GW$) 
provides a unique assessment and evaluation of ground-state properties 
unbiased by the starting-point dependence that characterizes the 
perturbative approaches.\cite{PhysRevB.86.081102,stan2,rubioscgw,hvb07,kressescrpa,gorlingscrpa} 
Moreover, self-consistency is essential to investigate the impact of 
advanced xc approximations on the electron density, 
since perturbative approaches do not alter the underlying wave function.\cite{PhysRevB.90.085141}

The total-energy curve of covalently-bonded diatomic molecules provides an 
important and difficult test case for both MBPT and DFT approximations. 
The dissociation of molecules with open-shell atoms such 
as H$_2$ and LiH are characterized by a large degree of static correlation, i.e., 
the electronic wave-function is not representable in terms of a single 
Slater determinant. Therefore, the accuracy achieved in the description of the dissociation region
reflects the capability of a given xc approximation to 
capture (or mimic) the multiple Slater determinant 
character of the wave function. It is already well-known that RPA 
yields a good description of the total energy in the dissociation region,\cite{fngb05,Gorling2011,rpamori,thygesen2014} 
whereas $GW$ overestimates it considerably.\cite{stan2,PhysRevLett.110.146403} 
In addition to the problem of static correlation, dissociation tests the 
ability of a functional to localize the electrons, important to accurately 
capture the abrupt change in the density upon atomization. How well $GW$ 
and RPA perform in this regard is still unknown and will be addressed here.

In this paper we present a thorough analysis of sc$GW$ and scRPA for 
ground-state properties of diatomic molecules at dissociation. 
We investigate the impact of locality and non-locality 
on the different energy contributions to the total energy. We also 
illustrate that a non-local treatment of exchange and correlation in the 
$GW$ approximation opens the gap between the highest-occupied and lowest-unoccupied molecular 
orbitals (HOMO and LUMO). 
Furthermore, the frequency dependence of the $GW$ self-energy is shown 
to significantly modify the ground-state of the system, leading, e.g., to a 
density matrix with fractional occupation numbers. Although $GW$ can capture important 
many body physics it reaches very slowly to the dissociation limit.
The local RPA potential, on the other hand, approaches the dissociation limit faster and thus yields 
a better description of the ground state energy in the dissociation region.

A convenient way to test the performance of an approximate functional 
with respect to static correlation and the ability to localize the electrons is 
to determine the so-called fractional spin and fractional charge errors. The fractional spin 
and fractional charge errors of the RPA have been studied previously.\cite{Hellgren/Rohr/2012,rpamori} 
In order to estimate the same errors in $GW$ we have used a more indirect approach. 
To determine an approximate fractional spin error, we compare 
spin-unrestricted dissociation with spin-restricted dissociation and in order to 
determine a fractional charge error we study the dipole moment of LiH 
as a function of nuclear separation. In this way, the errors are not determined in 
the dissociation limit but at large finite $R$, far beyond the point of atomization. 
Both $GW$ and RPA are affected by rather large 
fractional charge errors and hence suffer from an insufficient ability to accurately 
localize the electrons during dissociation. While RPA is free from fractional spin error 
at large interatomic distances,  $GW$ also  exhibits a rather large fractional spin error. 

Our first-principles calculations are complemented by an analytic derivation of 
explicit formulae for the Green's function 
and the correlation energy of a model 2-level H$_2$ molecule.
This allows us to investigate the limit of infinite interatomic 
separation which is not accessible by numerical studies.
The results indicate that $GW$ and RPA are very similar in the {\em dissociation limit} 
but that $GW$ is subjected to a very slowly converging gap leading to a very different 
behavior in the dissociation region.  The spurious local maximum characteristic of the 
RPA total energy curve could therefore also be present in $GW$, albeit largely extended. 

The paper is organized as follows.
In Sec.~\ref{sec:theory}, we introduce the $GW$ and RPA total 
energy functionals.
In Sec.~\ref{sec:etot}, we discuss the problem of covalent bond 
dissociation and the impact of non-locality and 
frequency dependence on the ground-state properties of the system.
Section~\ref{sec-2level} presents the derivation of an analytic expression for the 
one-shot $GW$ Green's function and for the $GW$ and RPA correlation energy 
in the dissociation limit for a simplified two-level model. 
We present an analysis of the fractional spin and fractional charge errors in MBPT 
in Sec.~\ref{sec:fractional}. Finally, our summary and conclusions are presented in 
Sec.~\ref{sec:conclusion}.

\section{Theory}\label{sec:theory}
The usual and most direct way of calculating the total energy 
from the single-particle Green's function $G$ is via the Galitskii-Migdal (GM) formula:\cite{galitskii}
\begin{align}
E_{\rm tot}[G] =T[G]+E_{\rm ext}[G]+E_{\rm H}[G]+E_{\rm xc}[G],
\label{eq:etot}
\end{align}
in which $T$ denotes the kinetic energy, $E_{\rm ext}$ the external potential energy, and $E_{\rm H}$ the Hartree 
energy. The exchange-correlation (xc) energy 
\be\label{eq:gm-xc}
E_{\rm xc}[G]=\int_0^\infty\! \frac{d\w}{2\pi}\Tr\!\{\S(i\w)G(i\w)\},
\ee
is determined from $\S$ which is here defined as the self-energy minus the Hartree potential $v_{\rm H}(\vr)=\int d\vr' n(\vr')v(\vr-\vr')$. The self-energy is also needed to 
compute $G$ via Dyson's equation
\be
G=G_{\rm H}+G_{\rm H}\S [G]G.
\label{dyson}
\ee
Notice that spatial and time coordinates have been suppressed in order to keep the notation light.  
The GM expression is non-variational 
(i.e., ${\delta E_{\rm tot}[G]}/{\delta G} \neq 0$ when $G$ is the solution of the Dyson equation). 
Therefore,  when evaluating Eq.~(\ref{eq:etot}),  it is necessary that $G$ is obtained from 
the iterative solution of Eq.~(\ref{dyson}) in order to get accurate results (see, e.g., Refs.~\onlinecite{PhysRevB.86.081102,stan}).
In MBPT it is also possible to formulate energy functionals which are 
variational with respect to $G$ (i.e. ${\delta E_{\rm tot}[G]}/{\delta G} = 0$ when $G$ satisfies Eq.~(\ref{dyson})). Several different kinds have been proposed\cite{luttingerward1960,Almbladh/etal:1999,barth05} 
but the most simple is the one introduced by Klein:\cite{klein1961}
\be
E_{\rm K}[G]= -i \F[G] +i \Tr\{GG_{\rm H}^{-1}-1 +\ln(-G^{-1})\}+E_{\rm H}.
\label{energi}
\ee
The $\Phi$-functional is related to the self-energy by 
$\S = \delta \Phi / \delta G$. It is easy to verify that in any approximate but 
$\Phi$-derivable self-energy $E_{\rm K}$ is stationary when $G$ 
obeys Dyson's equation.
Furthermore, at the stationary point $E_{\rm K}$ equals the total energy 
obtained from the corresponding GM formula. 
However, the Klein functional 
is more advantageous than the GM expression when considering approximate Green's functions 
since the variational property of the Klein functional ensures that the total energy will be close 
to that evaluated with the self-consistent $G$. 
Thus, the variational nature of the Klein functional can be employed to avoid the numerical complexity of the 
Dyson's equation, and at the same time obtain an energy close to the self-consistent one. 
The accuracy to which this can be achieved has been investigated in previous work\cite{stan2} and 
will be highlighted also in the present work. 

We will now specialize the discussion to two $\Phi$-derivable approximations: the 
Hartree-Fock/exact-exchange (HF/EXX) and the $GW$/RPA approximations. 
At the HF level the $\Phi$  functional assumes the following form: 
\be
\Phi=\frac{i}{2}\Tr{\{GGv\}}.
\label{phif}
\ee
It can be easily verified that applying the functional derivative $\Sigma=\delta\Phi/\delta G$ to Eq.~\ref{phif}  yields the 
Fock self-energy $iGv$, which is non-local and frequency independent.
The HF Green's function will thus take the form of a non-interacting Green's function. 
Evaluating Eq.~(\ref{energi}) with an arbitrary non-interacting Green's function -- here denoted $G_s$ -- results in\cite{barth05} 
\be
E_{\rm K}= T_s +  E_{\rm ext} + E_{\rm H}-i \F[G_s] ,
\label{energi2}
\ee
where $T_s$ is the kinetic energy of non-interacting electrons. When $\Phi$ is given by Eq.~(\ref{phif}) this is 
just the standard HF expression for the total energy. By comparing Eq.~(\ref{eq:etot}) and (\ref{energi2}), one sees 
that at the HF level and {\em at a non-interacting $G_s$}
the Klein expression and the GM formula coincide. 

We can further restrict the Green's function $G_s$ to come from a {\em local} potential. 
In this way Eq.~(\ref{energi2}) has the same form as the total energy in 
KS DFT and $-i \F[G_s]$ can be identified with the KS xc energy.\cite{barth05} 
If the HF approximation is constrained  to be evaluated with a $G_s$ from a local potential, 
the Klein functional reduces to the EXX functional.
It is well known that when the EXX functional is optimized 
with respect to the local EXX potential -- as opposed to the non-local 
HF self-energy -- it produces an energy very similar to (and slightly higher than) HF. 
For example, the difference in total energy of atoms is around 
10 parts per million.\cite{gkkg00} The HF Klein functional is 
thus quite stable with respect to variations in the Green's function. 

\begin{figure*}
\includegraphics[width=0.9\textwidth]{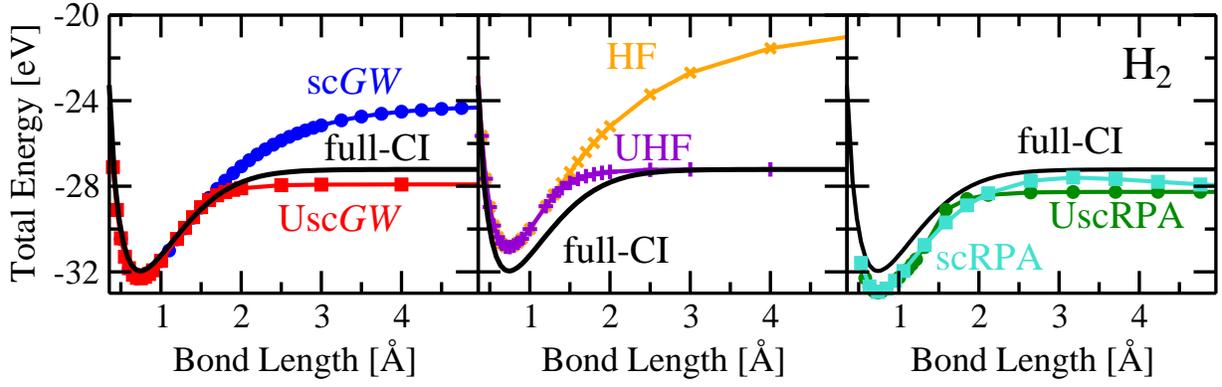}
\caption{\label{unrest} H$_2$ dissociation curves in spin-unrestricted/restricted self-consistent (Usc/sc) $GW$, RPA and HF.  Results are compared to accurate full-CI results from Ref. \onlinecite{h2ci1993}.}
\end{figure*}

Let us now turn to the $GW$ approximation for which
the self-energy is given by
\be
\S=iGW=iG[v+v\chi^{\rm H} v]
\label{gwself}
\ee
where 
\be
\chi^{\rm H}=\chi_0+\chi_0v\chi^{\rm H}
\label{chih}
\ee
is the reducible polarizability in the Hartree approximation and $\chi_0=iGG$ is the zeroth order approximation to the irreducible polarizability.
In the $GW$ approximation for $\Sigma$, the GM exchange-correlation energy in Eq.~\ref{eq:gm-xc} reduces to:
\be
E_{\rm xc}[G]=E_{\rm x}[G]+\int_0^\infty \frac{d\w}{2\pi}\Tr\{v[\chi^{\rm H}(i\w)-\chi_0(i\w)]\}.
\label{gwcorr}
\ee
Equations~(\ref{gwcorr}) and (\ref{eq:etot}) provide the explicit form of the GM energy in the $GW$ approximation. 
At self-consistency the GM energy is equal to that obtained from the Klein functional. Therefore, 
due to the simplicity of the GM expression and its natural decomposition into different energy 
contributions it is more convenient to use the GM expression whenever the self-consistent 
$GW$ Green's function is considered. All the self-consistent results presented in the next sections correspond to the 
evaluation of the GM formula.

Let us now evaluate the Klein $GW$ functional with a non-interacting $G_s$. 
The $\Phi$-functional generating the $GW$ self-energy is equal to
\be
\F=\frac{1}{2}\Tr\{\ln(1+ivGG)\}.
\label{phirpa}
\ee
It can be easily verified by functional differentiation of Eq.~(\ref{phirpa}) that $\Sigma=\delta\Phi/\delta G = iGW$.
If evaluated with a non-interacting Green's function $G_s$, Eq.~(\ref{phirpa}) takes the form of the RPA xc energy 
functional: 
\bea
\frac{1}{2}\Tr\{\ln(1+ivG_sG_s)\}&=&E_{\rm x}[G_s]+\nn\\
&&\!\!\!\!\!\!\!\!\!\!\!\!\!\!\!\!\!\!\!\!\!\!\!\!\!\!\!\!\!\!\!\!\!\!\!\!\!\!\!\!\!\!\!\!\!\!\!\!\!\!\!\!+
\int _0^\infty\!\frac{d\w}{2\pi}\int_0^1\! d\l\, \Tr\{v[\chi^{\rm H}_\l(i\w)-\chi_s(i\w)]\}
\label{rpacorr}
\eea
where $E_\x=\frac{i}{2}\Tr{G_sG_sv}$ is the Fock exchange energy, $\chi_s=iG_sG_s$ is the 
irreducible polarizability (or non-interacting density response function) evaluated with the KS Green's function 
and $\chi^{\rm H}_\l$ is the reducible polarizability (or the Hartree density response function) 
of a system with a linearly scaled Coulomb interaction $\l v$, i.e.
\be
\chi^{\rm H}_\l=\chi_s+\l\chi_s v\chi^{\rm H}_\l.
\ee
The RPA energy is therefore just the $GW$ Klein functional evaluated with a non-interacting 
Green's function. If we compare Eqs.~(\ref{eq:etot}) and (\ref{gwcorr}) to Eqs.~(\ref{energi2}) and (\ref{rpacorr}) we see that apart from the input Green's function the $GW$ and RPA energy expressions differ only by a coupling constant integral. 

To facilitate the comparison and make a clear distinction between the $GW$ and the RPA energies we 
will from now on denote the GM $GW$ correlation energy $U_{\rm c}^{GW}$ and the Klein RPA correlation energy 
$E_{\rm c}^{\rm RPA}$:
\begin{align}
U_{\rm c}^{GW}&=\int_0^\infty \frac{d\w}{2\pi}\Tr\{v[\chi^{\rm H}(i\w)-\chi_0(i\w)]\}\label{eq-uc}\\
E_{\rm c}^{\rm RPA}&=\int_0^\infty \frac{d\w}{2\pi}  \int_0^1\! d\l\,\Tr\{v[\chi_\lambda^{\rm H}(i\w)-\chi_s(i\w)]\}.
\end{align}
Differently from $U_{\rm c}$, $E_{\rm c}$ incorporates also the so called kinetic correlation energy -- defined as 
the difference between the kinetic energy of the interacting system and that of the fictitious non-interacting particle system -- 
which is included through the adiabatic-connection integrations over the interaction strength $\lambda$.
$U_{\rm c}$, on the other hand, includes only electronic correlation arising from the Coulomb interaction. 
It is therefore possible to define the kinetic correlation energy in the RPA through the following expression:
\be\label{eq-kincorr}
T_{\rm c}^{\rm RPA}[G_s]\equiv E^{\rm RPA}_{\rm c}[G_s]-U^{GW}_{\rm c}[G_s].
\ee
When the Klein and the GM functionals are evaluated with a non-interacting Green's function, 
the total energies obtained from the two approaches will differ exactly by 
$T_{\rm c}$, since the remaining energy terms in the Klein and GM formula have the same functional dependence on $G_s$. 
Using Eq.~(\ref{eq-kincorr}), we can rewrite the RPA total energy as
\bea\label{eq-rpaen}
E^{\rm RPA}[G_s]&=&T_s[G_s]+T^{\rm RPA}_c[G_s]+E_{\rm ext}[G_s]\nn\\
&&+E_{\rm H}[G_s]+E_{\rm x}[G_s]+U^{GW}_{\rm c}[G_s].
\eea
Notice again that this rewriting of the Klein functional as a functional of $G_s$ is only possible for a 
non-interacting $G_s$. The GM expression instead is defined for any $G$ 
and we will call this the $GW$ energy
\bea\label{eq-gwen}
E^{GW}[G]&=&T[G]+E_{\rm ext}[G]+E_{\rm H}[G]\nn\\
&&+E_{\rm x}[G]+U^{GW}_{\rm c}[G]
\eea
In the next section we will use Eqs.~(\ref{eq-rpaen}) and (\ref{eq-gwen}) for comparing the different energy contribution in the $GW$ and the RPA.
\begin{figure*}[t]
\includegraphics[angle=-90,width=1.0\textwidth]{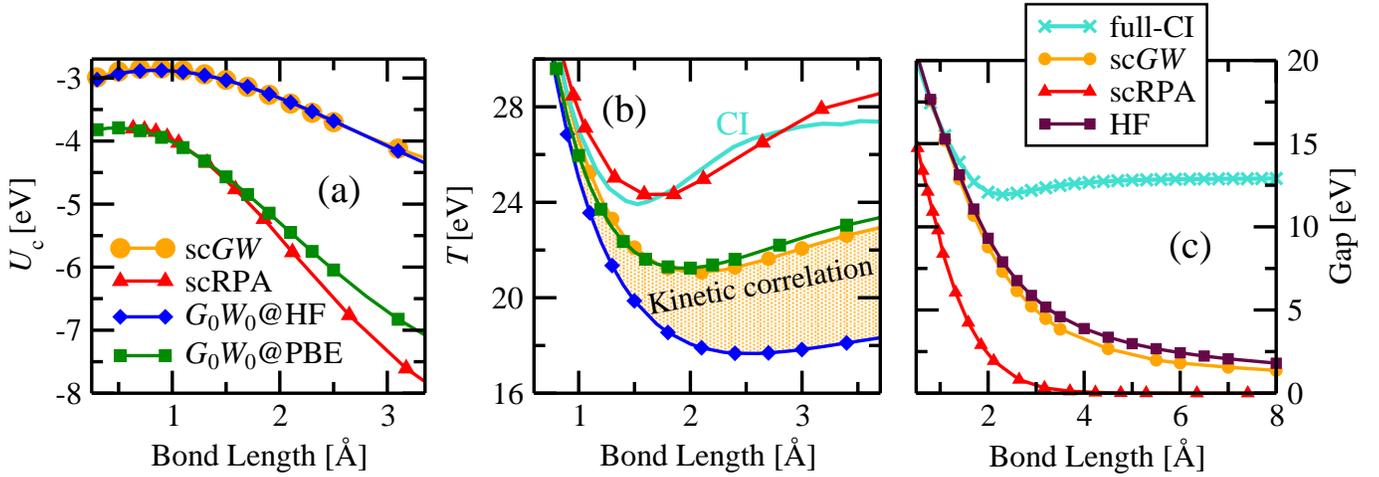}
\caption{\label{kinetic}(a) $U_{\rm c}^{GW}$ evaluated with the KS Green's function in PBE and RPA (denoted $G_0W_0$@PBE and scRPA) and with $GW$ and HF Green's functions (denoted $G_0W_0$@HF and sc$GW$). (b) Kinetic energies calculated with different Green's functions as $T[G^{GW}]$ for sc$GW$, $T[G^{\rm HF}]$ for $G_0W_0$@HF, $T[G^{\rm PBE}]$ for $G_0W_0$@PBE and $T[G^{\rm RPA}]+T_{\rm c}[G^{\rm RPA}]$ for scRPA. We also report the exact kinetic energy from the full-CI calculations of Ref.~\onlinecite{robert1996}. (c) Corresponding HOMO-LUMO gaps.}
\end{figure*}
In the following, we evaluate the Klein and GM functionals employing Green's functions evaluated 
in the RPA and $GW$ approximations. To avoid possible dependences on the starting point, both 
approaches are iterated to full self-consistency -- denoted as self-consistent $GW$ (sc$GW$) and 
self-consistent RPA (scRPA).
In sc$GW$ the Green's function is obtained from the iterative solution of 
Dyson's equation (Eq.~\ref{dyson}) with $\Sigma$ in the $GW$ approximation (Eq.~\ref{gwself}). 
The sc$GW$ method has been implemented in the FHI-aims code.\cite{blum,Ren/etal:2012} 
Details of the sc$GW$ implementation can be found elsewhere.\cite{PhysRevB.88.075105}
If not otherwise stated, coupled-cluster singles doubles (CCSD) and full configuration interaction (full-CI)
calculations have been performed with the ORCA code package\cite{WCMS:WCMS81} using Dunning's 
cc-PVDZ and cc-PVTZ Gaussian basis sets\cite{gaussianbasis1989} and extrapolated to the complete basis set limit.

In scRPA the energy is optimized with respect to a local potential determined by the optimized effective 
potential equation (also known as the Linearized Sham-Schl\"uter (LSS) equation).\cite{ShamSchluter,Casida95}
In RPA it is given by\cite{PhysRevA.68.032507,hvb07}
\begin{equation}
\label{eq:SSE}
\int \chi_sv_{\rm xc}^{\rm RPA}=-i\int \frac{d\w}{2\pi} \S^{GW}[G_s](\w)G_s(\w)G_s(\w)\
\end{equation}
where spacial variables have been suppressed. Details on the implementation can be found in 
Ref. \onlinecite{Hellgren/Rohr/2012}.

For comparison, we will also consider the GM and Klein functional evaluated using Green's functions from the 
HF approximation and the semi-local PBE approximation.\cite{PBE} In these cases, the corresponding GM (Klein) 
total energy functional is equivalent to the inclusion of $GW$ (RPA) 
correlation energy $U_{\rm c}^{GW}$ ($E_{\rm c}^{\rm RPA}$) 
at the level of first-order perturbation theory.
This allows us to establish a connection with previous works 
in which the RPA and $GW$ total energies have been considered 
at a perturbative level.

\section{Total energy}\label{sec:etot}
In this section, we present an analysis of the different contributions to the H$_2$ total 
energy curve. The results illustrate how a non-local and frequency dependent treatment of 
exchange and correlation effects may affect the ground-state properties of diatomic molecules
composed of open-shell atoms.
\begin{figure}[b]
\includegraphics[width=0.42\textwidth]{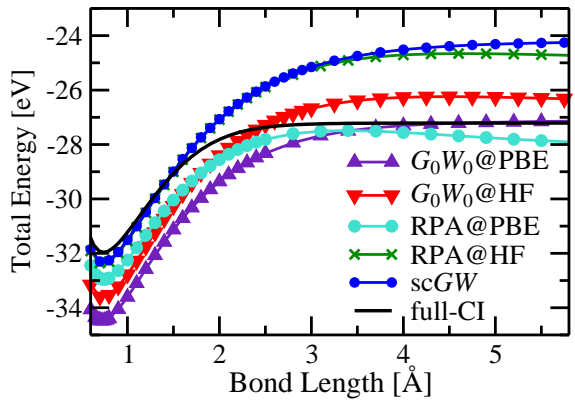}
\caption{\label{fig-perturbative} Total energy of H$_2$ evaluated in RPA and $GW$ based on first-order 
perturbation theory and full-CI results\cite{h2ci1993}.  
}
\end{figure}

\subsection{Coulomb energy and the HOMO-LUMO gap}
In Fig.~\ref{unrest} we report the dissociation curves of H$_2$ 
evaluated with sc$GW$, scRPA, and HF. 
Exact reference values from full-CI
calculations are included for comparison.\cite{h2ci1993} 
Due to the variational properties of the Klein functional, one may expect that
sc$GW$ and scRPA provide a qualitatively similar description of the total energy 
along the entire dissociation curve.
However, as observed in Ref.~\onlinecite{PhysRevLett.110.146403}, 
the sc$GW$ and scRPA total energies differ considerably for bond lengths larger than 2~\AA.

In the following, we illustrate that the Coulomb correlation energy 
depends significantly on whether it is evaluated with a 
Green's function derived from a local or a non-local potential.  
The total energy is decomposed according to Eqs.~\ref{eq-rpaen} 
and \ref{eq-gwen}, which allows us to analyze the different 
contributions separately using different Green's functions.
In the left panel of Fig.~\ref{kinetic} we report the Coulomb part of the 
correlation energy $U_c^{GW}[G]$ (given by Eq.~\ref{eq-uc}) evaluated with Green's 
functions obtained from PBE, scRPA, HF, and sc$GW$. 
The correlation energies evaluated from sc$GW$ and $G_0W_0$@HF are in very good agreement, 
and so are the $U_c^{GW}[G]$ derived from the scRPA and PBE Green's functions. 
However, the correlation energy is significantly larger when the Green's 
function is derived from a local potential (scRPA and PBE) as compared to a 
non-local one (sc$GW$ and HF).
In fact, the use of a local potential increases the Coulomb energy by almost a factor of 2 
in the dissociation region.
As analyzed in Ref.~\onlinecite{PhysRevLett.110.146403} 
this behaviour is intimately related to the HOMO-LUMO (or quasi-particle) gap contained in 
the Green's function. In the KS system of a molecule with open-shell atoms like the H$_2$ molecule 
the gap is largely underestimated and decays exponentially fast to zero with nuclear separation.  
A non-local potential is thus required to reproduce the true quasiparticle gap. Indeed, in the right 
panel of Fig.~\ref{kinetic} we see that the HF and sc$GW$ gaps are closer to the true gap, at least up 
to 2\,\AA. At larger $R$ the exact full-CI gap quickly reaches a finite value corresponding to the difference
between the ionization energy and the electron affinity of the Hydrogen atom. 
In contrast, all approximate methods considered here yield a vanishing HOMO-LUMO gap at dissociation. 
Furthermore, the HF and sc$GW$ approximations exhibit 
a very slow convergence (with respect to increasing 
bond length separation) to the dissociation limit.
Since $U_c^{GW}$ decays as the square root of the gap\cite{dahlenleeuwen2006,Gorling2011} 
the large difference in Coulomb correlation energy when using a local and a non-local 
potential is a direct reflection of this slow convergence. 
The interdependence of the HOMO-LUMO gap and the Coulomb correlation energy 
$U_c^{GW}$ is investigated in more detail in Sec.~\ref{sec-2level}, 
where the analytical solution of the Green's function for a two-level model of 
the H$_2$ molecule is presented. 

In Fig.~\ref{fig-perturbative} the RPA total energy is calculated with a HF Green's function (RPA@HF). 
The energy is now very close to the self-consistent $GW$ result showing that the non-locality 
of the self-energy is a crucial factor for the variational property of the Klein energy 
expression. The difference between RPA@HF and $G_0W_0$@HF is exactly $T_c[G^{\rm HF}]$ which
can be seen to be almost constant along the dissociation curve. In the next section we further analyze 
kinetic correlation effects. 

\subsection{Kinetic energy and density matrix analysis}
In the middle panel of Fig.~\ref{kinetic}, we report the kinetic contributions 
to the total energy evaluated in PBE, HF, sc$GW$, and scRPA. 
The scRPA kinetic energy, defined as the sum of the non-interacting kinetic energy
and the kinetic correlation energy (Eq.~\ref{eq-kincorr}), is in close agreement with 
exact full-CI results.~\cite{robert1996}

Just like the Coulomb correlation energy $U_{\rm c}$, also 
the kinetic energy depends strongly on the input Green's function.
However, the sc$GW$ ($T[G^{GW}]$) and $G_0W_0$@PBE ($T[G^{\rm PBE}]$) 
kinetic energies assume similar values at every $R$. 
When evaluated with a HF Green function  $T[G^{\rm HF}]$, on the other hand, the 
kinetic energy decreases by approximately 15-20\%.
This underestimation can explain almost the most of the difference between sc$GW$ and 
$G_0W_0$@HF around equilibrium (see Fig.~\ref{fig-perturbative}).
However, when stretched, a difference in the external, Hartree and exchange energies 
also become important indicating that the difference in the density and the density matrix
between HF and sc$GW$ increases with separation.
\begin{figure}[t]
\includegraphics[width=0.45\textwidth]{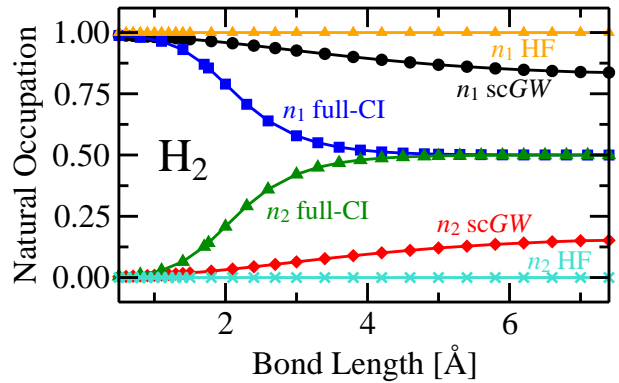}
\caption{\label{nat_gap} Natural occupation numbers obtained by diagonalizing the density matrix.}
\end{figure}
Indeed, in addition to the importance of screening the Coulomb potential, 
a qualitative change in the $GW$ ground-state at self-consistency can occur due the frequency 
dependence of the $GW$ self-energy. 
Figure~\ref{nat_gap} shows the first two natural occupation numbers ($n_1$ and $n_2$) of the 
density matrix -- i.e., the eigenvalues obtained from the density-matrix 
diagonalization -- in the different approximations compared to exact reference data 
from full-CI calculations. 
As expected, HF behaves as an independent particle approximation, i.e. 
the density matrix occupation numbers are integer at all bond lengths  -- reflecting
the fact that the density matrix is idempotent for non-interacting particle systems.
Identical results are obtained for PBE and scRPA.
On the other hand, according to exact full-CI calculations 
$n_1$ and $n_2$ should both approach $0.5$ in the dissociation limit, 
due to the multiple Slater-determinant character of the electronic wave function 
at dissociation. 
The $GW$ density matrix yields fractional occupation numbers at all bond lengths. 
Therefore, the dynamical $GW$ self-energy leads to 
interacting ground-state properties that are qualitatively different 
from the non-interacting particle ground-state of HF, PBE, and scRPA. 
However, the occupation numbers still largely deviate from the full-CI values,
and at 6~\AA \, $n_1$ and $n_2$ are far from the exact results. 
In the next section, we illustrate that for a two-level model for H$_2$, 
the occupation $n_1$ and $n_2$  reach full degeneracy ($n_1=n_2=0.5$), 
at dissociation.
However, this limit is only reached at very large interatomic separation 
due to the slow variation of the occupation numbers ($\sim \D\ve^{1/4}$) on the 
HOMO-LUMO gap $\D\ve$. This slow convergence to the dissociation limit, 
further discussed in Sec. IV, appears a characteristic feature of the $GW$ approximation. 

\section{Dissociation limit of H$_2$ in a minimal basis}\label{sec-2level}
To gain further insight into the dissociation of H$_2$ and, in particular, the dissociation limit ($R\to\infty$), 
in $GW$ and RPA we will study a minimal basis H$_2$ 
model system described by the HOMO and the LUMO orbitals only. 
The same model has previously been used 
to study the dissociation limit of the 
RPA\cite{fngb05}and the quasiparticle gap in the perturbative $GW$ approximation.\cite{ask12} More 
recently, the similar two-site Hubbard model was used to capture qualitative features of 
H$_2$-dissociation for approximations beyond the RPA.\cite{thygesen2014}
\subsection{Minimal basis model for H$_2$}
At large nuclear separation $R$ the HOMO is approximated by the symmetric linear combination of the 
atomic orbitals, localized at atom $A$ and $B$, 
\be
\vf_H(\vr)=\frac{1}{\sqrt{2}}\left[\vf_A(\vr)+\vf_B(\vr)\right],
\label{homo}
\ee
and the LUMO is approximated by the corresponding antisymmetric linear combination
\be
\vf_L(\vr)=\frac{1}{\sqrt{2}}\left[\vf_A(\vr)-\vf_B(\vr)\right].
\label{lumo}
\ee
We can then construct the non-interacting (or KS) time-ordered Green's function 
\be
G_s(\vr,\vr',\w)=\frac{\vf_H(\vr)\vf_H(\vr')}{\w-\ve_H-i\eta}+\frac{\vf_L(\vr)\vf_L(\vr')}{\w-\ve_L+i\eta},
\label{gshl}
\ee
where $\ve_{H/L}$ is the HOMO/LUMO eigenvalue.
The interacting model Green's function is determined by inverting the Dyson equation (Eq.~(\ref{dyson})) 
\bea
G(\vr,\vr',\w)&=&\sum_{ij}\vf_i(\vr)[A^{-1}(\w)-\S(\w)]^{-1}_{ij}\vf_j(\vr').
\label{dysmod}
\eea
The sum runs over the HOMO and the LUMO and the matrix $A$ is equal to
\be
A^{-1}=\left[\begin{array}{cc}\w-\ve_H-i\eta & 0 \\0 & \w-\ve_L+i\eta\end{array}\right].
\ee
A matrix element of the self-energy is determined from $\S_{ij}(\w)=\int d\vr_1d\vr_2\vf_i(\vr_1)\S(\vr_1,\vr_2;\w)\vf_j(\vr_2)$. 
In the HF approximation the self-energy matrix becomes diagonal 
\be
\S=\left[\begin{array}{cc}\frac{1}{2}[U_0+U_1] & 0 \\0 & \frac{1}{2}[U_0+3U_1]\end{array}\right],
\label{sigmahf}
\ee
with
\bea
U_0&=&\bra \vf^2_A|v|\vf^2_A\ket=\bra \vf^2_B|v|\vf^2_B\ket\\
U_1&=&\bra \vf^2_B|v|\vf^2_A\ket.
\eea
Making use of Eq.~(\ref{sigmahf}), it is straightforward to evaluate Eq.~(\ref{dysmod}). We find
\bea
G_{HF}(\vr,\vr',\w)&=&\frac{\vf_H(\vr)\vf_H(\vr')}{\w-\ve_H-\frac{1}{2}U_0-\frac{1}{2}U_1-i\eta}\nn\\
&&+\frac{\vf_L(\vr)\vf_L(\vr')}{\w-\ve_L-\frac{1}{2}U_0-\frac{3}{2}U_1+i\eta}.
\eea
The quasiparticle gap thus increases with $U_1$ in the HF approximation. As $R\to \infty$, $U_0$ remains finite
while $U_1$ vanishes as $1/R$.  The HF gap therefore reduces to the non-interacting HOMO-LUMO gap in the 
dissociation limit but decays as $1/R$ which can be compared to the exponentially fast decay of the non-interacting gap.

\subsection{\textit{GW} gap in the dissociation limit}
Let us now turn to the $GW$ approximation and evaluate the self-energy with Eq.~(\ref{gshl}). Hence, $\S=iG_s[v+v\chi v]$ and
\be
\chi=\chi_s+\chi_sv\chi
\label{rpa}
\ee 
with $\chi_s=-iG_sG_s$. In terms of orbitals and eigenvalues we can write
\be
\chi_s=2\left[\frac{f_s(\vr)f_s(\vr')}{\w-\D\ve+i\eta}-\frac{f_s(\vr)f_s(\vr')}{\w+\D\ve-i\eta}\right]
\label{chis}
\ee
where $f_s(\vr)=\vf_H(\vr)\vf_L(\vr)$ and $\D\ve=\ve_L-\ve_H$. Inserting Eq.~(\ref{chis}) into Eq.~(\ref{rpa}) 
we find 
\be
\chi=2\frac{\D\ve}{\D E}\left[\frac{f_s(\vr)f_s(\vr')}{\w-\D E+i\eta}-\frac{f_s(\vr)f_s(\vr')}{\w+\D E-i\eta}\right]
\label{chirpa}
\ee
where $\D E=\sqrt{\D \ve^2+4v_{s}\D\ve}$ and $v_s=\bra f_s|v|f_s\ket$. 

The correlation part of the self-energy in the $\vr$ and $\vr'$ basis is then easily determined
\bea
\S_c&=&i\int \frac{d\w'}{2\pi}G_s(\w+\w')v\chi(\w') v=\nn\\
&&\!\!\!\!\!\!\!\!\!\!\!\!\!\!\!2\frac{\D\ve}{\D E}\left[\frac{\l_{HH}(\vr',\vr)}{\w+\D E-\ve_H-i\eta}+\frac{\l_{LL}(\vr',\vr)}{\w-\D E-\ve_L+i\eta}\right]
\eea
in which
\bea
\!\!\!\!\!\!\l_{kk}(\vr',\vr)&=&\nn\\
&&\!\!\!\!\!\!\!\!\!\!\!\!\!\!\!\!\!\!\!\!\!\!\!\!\!\!\!\!\!\!\!\!\!\!\!\!\int d\vr_1d\vr_2\vf_k(\vr)v(\vr,\vr_1)f_s(\vr_1)f_s(\vr_2)v(\vr_2,\vr')\vf_k(\vr').
\eea
Alternatively, one may express the self-energy as a matrix in $\vf_H$ and $\vf_L$
\be
\S_c=\left(\begin{array}{cc}\frac{\D\ve}{2\D E}\frac{(U_0-U_1)^2}{\w-\D E-\ve_L+i\eta} & 0 \\0 & \frac{\D\ve}{2\D E}\frac{(U_0-U_1)^2}{\w+\D E-\ve_H-i\eta}\end{array}\right)
\ee
which is diagonal. In order to solve the Dyson equation with this self-energy we first notice that we can write
\be
G=G_{\rm HF}+G_{\rm HF}[\S_c +\D\S_{\rm HF}]G,
\label{deltahf}
\ee
where $\D\S_{HF}$ is the difference between the HF self-energy evaluated with a $GW$ and a HF Green's function. 
In the following we will mainly consider the one-shot solution for which this 
term is zero. In general this term is not expected to contribute qualitatively. The $GW$ self-energy is frequency dependent which complicates the extraction of the new poles. In Ref. \onlinecite{ask12} a perturbative solution was found. However, we can also solve the fully non-linear equation 
and after a few manipulations we find in total four solutions 
\bea
\!\!\!\!\!\!\!\!\Omega_\pm^{H}&=&\frac{\ve^0_H+\ve_L+\D E}{2}\pm \sqrt{\frac{(\D E+\ve_L-\ve^0_H)^2}{4}+k}\nn\\
\!\!\!\!\!\!\!\!\Omega_\pm^{L}&=&\frac{\ve_H+\ve^0_L-\D E}{2}\pm \sqrt{\frac{(\D E+\ve^0_L-\ve_H)^2}{4}+k}
\label{exc}
\eea
where we have defined
\be
k=\frac{\D\ve}{2\D E}(U_0-U_1)^2.
\ee
In order to distinguish between the Green's function inserted in the self-energy and the zeroth order Green's function, chosen to be that of the HF in Eq. (\ref{deltahf}),
we have denoted the eigenvalues of the latter $\ve^0_{H/L}$ and those of the former $\ve_{H/L}$. Both have the non-interacting form of 
Eq. (\ref{sigmahf}). 

The HOMO and LUMO are easily identified and we find the $GW$ gap 
\bea
\D&=&\Omega_+^{L}-\Omega_-^{H}=\frac{\ve_H+\ve^0_L-\ve^0_H-\ve_L}{2}-\D E\nn\\
&&+\sqrt{\frac{(\D E+\ve^0_L-\ve_H)^2}{4}+k}\nn\\
&&+\sqrt{\frac{(\D E+\ve_L-\ve^0_H)^2}{4}+k}.
\label{gap}
\eea
The two additional poles correspond to many-body excitations, i.e., they are accessible only when the 
ground state is described by more than one Slater determinant. In the two-level model one of these 
satellite excitations corresponds to the first excited state of the one-electron system and the other corresponds 
to the first excited state of the three-electron system.

\begin{figure}
\includegraphics[width=0.48\textwidth]{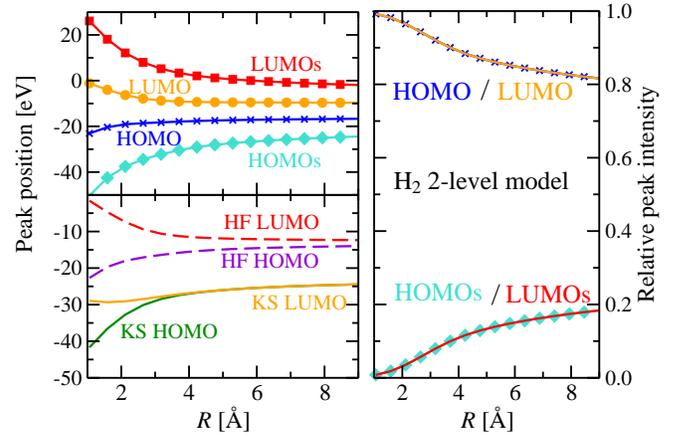}
\caption{\label{fig:gw-rpa_gap} Left panel: The four solutions (Eq. (\ref{exc})) of quasiparticle excitation energies in the one-shot $GW$ approximation (upper panel) and in the HF and KS-LDA  approximations (lower panel) for H$_2$  in a minimal basis. Right panel: Quasiparticle amplitudes of the four one-shot $GW$ solutions. }
\end{figure}

If the HF Green's function is plugged into the self-energy we obtain the one-shot $G_0W_0$@HF solution to the gap
\be
\D=-\D E+2\sqrt{\frac{(\D E+\D\ve)^2}{4}+k}.
\label{gaposhf}
\ee
In order to explicitly see the correction to the HF gap $\D\ve$ we can also write 
Eq.~(\ref{gaposhf}) as
\bea
\!\!\!\!\!\!\!\!\D&=&\D\ve\nn\\
&&\!\!\!\!\!\!\!\!\!\!\!\!+(\D E+\D\ve)\left[\sqrt{1+2\frac{\D\ve}{\D E}\frac{(U_0-U_1)^2}{(\D E+\D\ve)^2}}-1\right]
\label{gaposhf2}
\eea
from which we can conclude that the correction is always positive. Furthermore, expanding the square root 
we recover the perturbative result of Ref. \onlinecite{ask12} 
\be
\D\approx\D\ve+\frac{\D\ve}{\D E}\frac{(U_0-U_1)^2}{\D E+\D\ve}.
\label{gaposhf3}
\ee
In Fig.~\ref{fig:gw-rpa_gap} (left upper panel) we have plotted the four solutions in the one-shot $G_0W_0$@HF approximation 
as a function of $R$. The satellite excitations are denoted with an 's'. According to Eq. (\ref{exc}) all four solutions will 
converge to the same value in the dissociation limit. The $G_0W_0$@HF gap will thus also decay to zero but much slower 
as compared to HF (left lower panel). This can be attributed to the slowly decaying $1/\sqrt{R}$ terms that can be extracted 
from Eq.~(\ref{gaposhf}). We again stress that the exact gap should be finite in the 
dissociation limit (see Fig. \ref{kinetic}).

In order see what is the effect of self consistency we first rewrite Eq.~(\ref{gap}) as
\bea
\D&=&\frac{\D\ve-\D}{2}-\D E\nn\\
&&+\sqrt{\frac{(\D E+\D+\ve^0_L-\ve_L)^2}{4}+k}\nn\\
&&+\sqrt{\frac{(\D E+\D+\ve_H-\ve^0_H)^2}{4}+k}.
\eea
As a first step $\ve^0_L-\ve_L$ and $\ve_H-\ve^0_H$ are set to zero. In this way it is possible to find a unique solution 
for $\D$ which increases the gap even further. By instead approximating 
\be
\ve^0_L-\ve_L=\ve_H-\ve^0_H=\frac{\D E+\D}{2}-\sqrt{\frac{(\D E+\D)^2}{4}+k}
\ee
the gap is similar to the one-shot result. Proceeding to self-consistency closes the gap further. This result is consistent 
with the full basis result in Fig.~\ref{kinetic}. 

Let us now determine the quasiparticle amplitudes corresponding to the excitations in Eq.~(\ref{exc}). We find
\bea
Z_{\pm}^{H}&=&\left[\frac{1}{2}\mp\frac{\D E+\ve_{L}-\ve_{H}^0}{\sqrt{\frac{(\D E+\ve_{L}-\ve_{H}^0)^2}{4}+k}}\right]\vf_H(\vr)\vf_H(\vr')\\
Z_{\pm}^{L}&=&\left[\frac{1}{2}\pm\frac{\D E+\ve_{L}^0-\ve_{H}}{\sqrt{\frac{(\D E+\ve_{L}^0-\ve_{H})^2}{4}+k}}\right]\vf_L(\vr)\vf_L(\vr')
\eea
In the one-shot $G_0W_0$@HF approximation these reduce to
\bea
Z_{\pm}^{H}&=&\left[\frac{1}{2}\mp P\right]\vf_H(\vr)\vf_H(\vr')\\
Z_{\pm}^{L}&=&\left[\frac{1}{2}\pm P\right]\vf_L(\vr)\vf_L(\vr')
\eea
where 
\be
P=\frac{1}{4}\frac{\D E+\D\ve}{\sqrt{\frac{(\D E+\D\ve)^2}{4}+k}}.
\ee
The weight of the HOMO and LUMO excitations thus decreases with nuclear separation whereas the weights of the 
many-body excitations increases (see right panel of Fig. \ref{fig:gw-rpa_gap}). It is easy to see that $P$ tends to zero as $\D\ve^{1/4}$ with nuclear separation and thus 
all four excitations become degenerate and acquire the same weight of 1/2 in the dissociation limit. This means that the 
density matrix is correctly described in terms of natural occupation numbers in the dissociation limit.  However, due to the slow $\D\ve^{1/4}\propto R^{-1/4}$ decay the dissociation limit is reached only at very large $R$. The same behavior is found in the full basis results in Fig.~\ref{nat_gap}. 

The time-ordered one-shot $GW$ Green's function can finally be summarized as
\bea
\!\!\!\!\!\!\!\!\!\!\!\!\!\!G_{GW}&=&Z_{-}^{H}\frac{\vf_H(\vr)\vf_H(\vr')}{\w-\Omega_-^{H}-i\eta}+Z_{+}^{H}\frac{\vf_H(\vr)\vf_H(\vr')}{\w-\Omega_+^{H}+i\eta}\nn\\
&&+Z_{-}^{L}\frac{\vf_L(\vr)\vf_L(\vr')}{\w-\Omega_-^{L}-i\eta}+Z_{+}^{L}\frac{\vf_L(\vr)\vf_L(\vr')}{\w-\Omega_+^{L}+i\eta}.
\label{gwg}
\eea
\subsection{\textit{GW} and RPA correlation energy at dissociation}
We will now use the HF/KS and the one-shot $GW$ Green's functions to study the correlation energy in the dissociation limit. 
To this purpose we need to construct the zeroth order polarization propagator. For a HF/KS Green's function this was done in 
Eq.~(\ref{chis}). With the many-body $GW$ Green's function of Eq.~(\ref{gwg}) we find in total four poles with energies and oscillator strengths:
\bea
&&1: \D\ve_1=\Omega_+^{H}-\Omega_-^{H},\,\,\,f_1(\vr)=\sqrt{\frac{1}{4}-P^2}\,\vf_H(\vr)\vf_H(\vr)\nn\\
&&2: \D\ve_2=\Omega_+^{L}-\Omega_-^{L},\,\,\,f_2(\vr)=\sqrt{\frac{1}{4}-P^2}\,\vf_L(\vr)\vf_L(\vr)\nn\\
&&3: \D\ve_3=\Omega_+^{H}-\Omega_-^{L},\,\,\,f_3(\vr)=\left[\frac{1}{2}-P\right]\vf_L(\vr)\vf_H(\vr)\nn\\
&&4: \D\ve_4=\Omega_+^{L}-\Omega_-^{H},\,\,\,f_4(\vr)=\left[\frac{1}{2}+P\right]\vf_L(\vr)\vf_H(\vr).\nn\\
\eea
Both the correlation energy in RPA and the correlation energy in sc$GW$ contain the term $-\int \Tr\{v\chi_0\}$.
It is easy to see that this term gives identical results in the dissociation limit independently of which of the above $G$ is 
used for constructing $\chi_0$. In both cases we find
\be
-\lim_{R\to \infty}\int _0^\infty\frac{d \w}{2\pi}\Tr\{v\chi_0(i\w)\}=-\frac{U_0}{2}.
\ee
This is exactly equal to the static correlation error of the EXX/HF functional ($U_0/2$) and hence this term 
cancels this error in both RPA and $GW$. We will now show that the second term of the correlation energy 
$-\int \Tr\{v\chi_H\}$ gives zero contribution in the dissociation limit. The interacting polarization propagator 
within HF/KS is given by Eq.~(\ref{chirpa}) and it therefore decays as $\sqrt{\D\ve}$ as $R\to \infty$. 
Performing the $\l$-integral does not change this result. This explains the difference in rate at which the 
asymptotic limit is reached by inserting HF or a KS Green's function. 

Now let's see what we find when we use the one-shot $GW$ Green's function.
Inserting this response function into the RPA equations leads to a $4\times 4$ matrix. We notice, however, that 
this matrix can be written in terms of two sub-blocks, one in terms of the first and second excitations and 
the second in terms of the third and fourth excitations. We thus have to diagonalize two $2\times 2$ matrices 
independently. For example, in terms of excitation 3 and 4 we find
\be
\left(\begin{array}{cc}\D\ve_3^2+v_{33} & v_{34} \\v_{43} & \D\ve_4^2+v_{44}\end{array}\right)
\ee
where
\be
v_{ij}=\bra \tilde{f}_i |v| \tilde{f}_j\ket, \,\,\tilde{f}_j=2\sqrt{\D\ve_j} f_j.
\ee
The eigenvalues $\l_\pm$ and eigenvectors $(d_1^\pm\,d_2^\pm)$ of this matrix are
\bea
\l_\pm&=&\frac{\D\ve_3^2+v_{33}+\D\ve_4^2+v_{44}}{2}\nn\\
&&\pm\sqrt{\frac{(\D\ve_3^2+v_{33}-\D\ve_4^2-v_{44})^2}{4}+v_{34}^2}
\eea
\bea
d_1^+&=&\frac{v_{34}}{\sqrt{v_{34}^2+(\l_+-\D\ve_3^2-v_{33})^2}}\nn\\
d_2^+&=&\frac{\l_+-\D\ve_3^2-v_{33}}{\sqrt{v_{34}^2+(\l_+-\D\ve_3^2-v_{33})^2}}\nn\\
d_1^-&=&\frac{v_{34}}{\sqrt{v_{34}^2+(\l_--\D\ve_3^2-v_{33})^2}}\nn\\
d_2^-&=&\frac{\l_--\D\ve_3^2-v_{33}}{\sqrt{v_{34}^2+(\l_--\D\ve_3^2-v_{33})^2}}\nn
\label{chi0}
\eea
which yields the new oscillator strengths
\be
\tilde f_3=2d^+_1\sqrt{\frac{\D\ve_3}{\sqrt{\l_+}}} f_3+2d^+_2\sqrt{\frac{\D\ve_4}{\sqrt{\l_+}}} f_4
\ee
\be
\tilde f_4=2d^-_1\sqrt{\frac{\D\ve_3}{\sqrt{\l_-}}}f_3+2d^-_2\sqrt{\frac{\D\ve_4}{\sqrt{\l_-}}} f_4
\ee
It is easy to see that also these tend to zero in the dissociation limit. Thus, even when inserting
a one-shot $GW$ Green's function the term $-\int \Tr\{v\chi_H\}$ will tend to zero in the 
dissociation limit, suggesting that the energy will be similar to that found in RPA or RPA@HF. 
From the full H$_2$ $GW$ dissociation curves we can thus not exclude the 
possibility that the $GW$ curve exhibits a very extended local maximum (i.e., a {\it bump}) that decays to the 
correct dissociation limit a very large interatomic separation. 
We will discuss this topic more in the next section.

\section{Fractional spin and fractional charge errors}\label{sec:fractional}

The fractional spin and fractional charge errors have been rigorously defined in 
Ref. \onlinecite{Cohen08082008}. 
The study of these errors consists in treating the
atoms in the dissociation limit independently but by means of ensembles that allow 
for fractional charge and spin. 
If the density that minimizes the energy is integrated to a fractional number of 
particles on the individual atoms, the functional suffers a fractional charge or delocalization error and 
if an atom with fractional spin has a different energy from an atom with integer spin the functional suffers a fractional 
spin or static correlation error. 

Within DFT and RPA a generalization in terms of ensembles 
was made in Ref. \onlinecite{Hellgren/Rohr/2012}. However, in $GW$, and 
MBPT, such a procedure is less straightforward.\cite{bruneval} 
In this work we have therefore used a more indirect approach to estimate fractional charge 
and fractional spin errors in $GW$ based on the calculation  of spin-unrestricted dissociation curves 
and dipole moments of hetero-atomic dimers.
\subsection{Spin-unrestricted dissociation}
We have seen that there is a fundamental difference between RPA and $GW$ in how 
they deal with static correlation at large or intermediate distances. RPA is almost 
free from static correlation error and describes correctly the dissociation of covalent bonds. 
On the other hand, sc$GW$ overestimate the total energy at dissociation, 
with an error comparable to the PBE functional.
It is well known that by breaking the spin symmetry it 
is possible to simulate charge localization and avoid static correlation errors 
in near degeneracy situations. However, despite the improved description of the 
total energy, the spin-unrestricted ground-state is unphysical since it has the wrong spin-density.

The spin symmetry breaking is achieved by localizing the spin-up and spin-down electrons 
on different atoms in the initialization of the calculation. 
For bond lengths close to the equilibrium one, this procedure yields identical wave functions and total 
energies as for spin-restricted calculations. 
However, for larger bond lengths the total energy obtained from unrestricted calculations 
lies at a lower energy than the restricted one. 

In Fig.~\ref{unrest} we report spin-unrestricted self-consistent (Usc) calculations of the 
H$_2$-dissociation curve in RPA, $GW$ and HF. 
Figure~\ref{unrest} illustrates that UscRPA is similar to spin-restricted scRPA across the 
entire dissociation region - consistent with the fact that RPA has no fractional spin error.\cite{Hellgren/Rohr/2012} 
The lack of fractional spin error ensures that the spin-restricted and unrestricted curves converge to the same 
value asymptotically. 

On the contrary, the Usc$GW$ dissociation curve differs considerably
from the spin-restricted sc$GW$ curve at large internuclear distances and improves the 
dissociation region as expected. This picture confirms that at finite but large $R$ (i.e., at 
bond lengths significantly larger than the point of atomization in the exact treatment), sc$GW$ 
suffers large fractional spin error. However, when inserting a 
HF Green's function in the $GW$ (Klein) energy expression a spurious maximum around 4 \AA  \,
can be observed (see Fig.~\ref{fig-perturbative}), indicating that the energy eventually may improve - albeit at very large separation.
These results are consistent with the two-level model studied in Sec.~\ref{sec-2level},
where we show that the $GW$ energy evaluated with $G_{\rm HF}$ converges 
to the RPA energy (i.e., to the correct dissociation energy) in the limit of large separation 
but much slower than in RPA due to the slow decay of the HF quasiparticle gap.
It is further shown that inserting a one-shot $GW$ 
Green's function does not change this result, which suggests that
the sc$GW$ total energy might present an  
extended local maximum (or {\it bump}) which decreases 
to the exact dissociation limit for $R\rightarrow \infty$. 
This bump is a well known feature of the RPA functional\cite{fngb05} 
and also in RPA it does not disappear with self-consistency.\cite{PhysRevLett.110.146403} 
From Fig.~\ref{unrest} we can further conclude that spin-unrestricted $GW$ and RPA calculations do not 
exhibit spurious maxima. 
\begin{figure}
\includegraphics[width=0.45\textwidth]{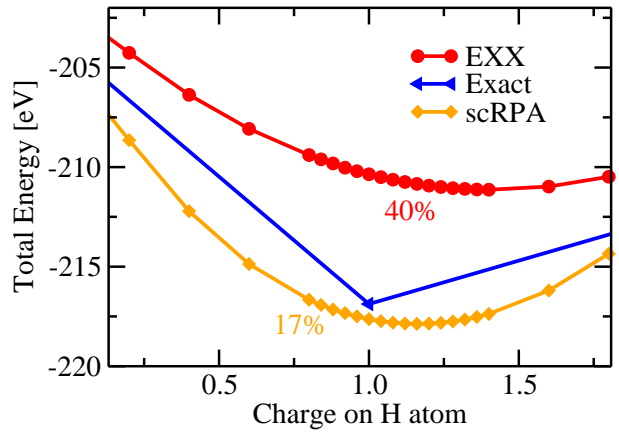}
\caption{\label{frac}Fractional charge analysis of the LiH molecule. The total energy as a function of fractional charge 
on the H atom. These curves are obtained by adding the fractional charge curves of the independent atoms in such a way that the total number of electrons is four.}
\end{figure}
\begin{figure*}
\includegraphics[angle=-90,width=0.9\textwidth]{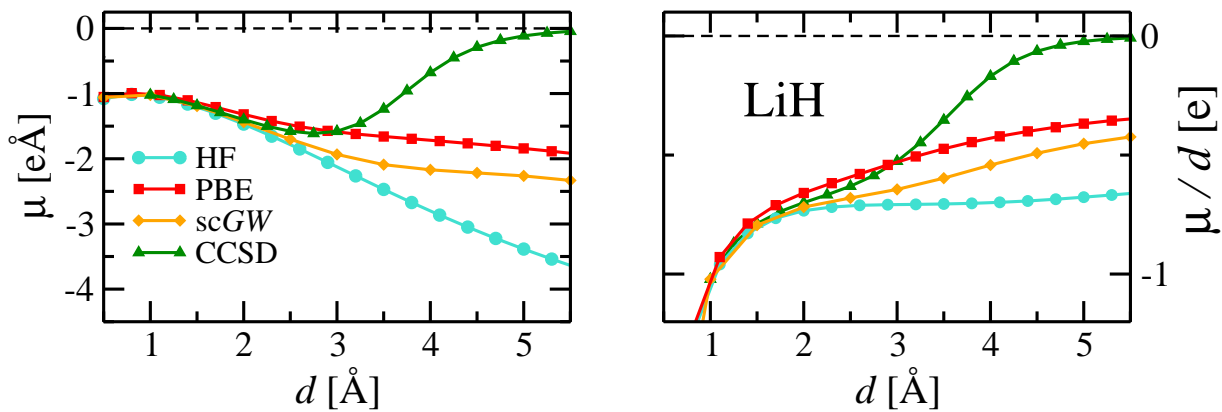}
\caption{\label{dipole}Left panel: The dipole moment of the LiH molecule as a function of bond distance in HF, PBE, sc$GW$, and accurate CCSD calculations. 
Right panel: The dipole moment divided by distance in the same approximations gives an estimate of their fractional charge error.}
\end{figure*}

\subsection{Dipole moment of LiH}
The problem of delocalization or fractional charge error is related to the tendency of approximate functionals 
to spread out, i.e. delocalize, the electrons in the system due to an insufficient treatment of the Coulomb repulsion. 
This problem gives rise to large errors when looking at response 
properties,\cite{vangis1999,kummel2004} but also results in an error in the total energy due to an incorrect electron 
distribution in the ground state. 

A very simple system which clearly reveals the delocalization error is the stretched LiH molecule. 
It is a heteronuclear system and therefore very sensitive to the ability of the functional to localize 
the electrons in the dissociation process. Molecules should dissociate into neutral atoms. However,
many approximations produce fractionally charge atoms in the dissociation limit of heteronuclear systems 
composed of open shell atoms\cite{Hellgren/Rohr/2012} and the amount of fractional charge can be seen 
as a measure of the delocalization error.
Obviously, homonuclear systems do not exhibit this behavior in the dissociation limit due to symmetry. 
This does, however, not mean that there is no delocalization problem. Also at finite separation the 
electrons can be too delocalized resulting in e.g. a slowly converging potential energy curve.  

A way to determine this error without actually calculating the energy 
of the infinitely stretched molecule is to do a fractional charge analysis. In DFT the functional is extended to ensemble 
densities that allows for non-integer number of electrons. The energy is then calculated as a function of particle number. 
Whereas the exact functional exhibits cusps at the integers and is linear between the integers most approximations 
produces smooth curves with a large curvature. To calculate the energy of the molecule at infinite separation
the atomic curves are added and the minimum is located. The exact functional minimizes non-analytically 
at the integer which ensures neutral dissociation. However, the usually smooth approximate curves lead to a
shift in the minimum as we will see below.

The generalization of the RPA to ensemble densities was done in Ref. \onlinecite{Hellgren/Rohr/2012}. In Fig.~\ref{frac} we 
have plotted the energy as a function of the total (fractional) number of electrons on the H atom. 
We see that EXX minimizes smoothly at around 1.4 electrons due to the non-linear behavior of the EXX functional. 
RPA is seen to reduce this error to 17\% but we still see a rather large curvature in the RPA.

In MBPT it is much more difficult to disentangle the dissociated atoms and treat them as fractionally charged isolated 
systems due to the frequency dependent self-energy. To circumvent this complication we have studied the problem by 
looking directly at the density of the stretched molecule via the dipole moment. In Fig.~\ref{dipole} we compare $GW$, 
HF, and PBE. 
For comparison, we also report coupled-cluster singles doubles 
(CCSD) dipole moments\cite{Szabo/Ostlund:1989} at the complete basis-set limit
performed with ORCA code package\cite{WCMS:WCMS81}.
At around 3.5~\AA\, the CCSD dipole moment exhibits a rather sharp transition from a 
finite value to zero. This indicates the breaking of the molecular bond into two neutral atoms. 
None of the approximations 
studied here is able to capture this transition. 
Instead the dipole moment steadily increases with separation due to the 
delocalization error. 
By dividing the dipole moment with the distance we estimate the fractional charge error at 
large (but {\em finite}) separation. 
At 6~\AA\, we find that the residual charge left on the 
Li/H fragments at dissociation amounts to 20\% of an electron 
in sc$GW$, 35\% in HF and 18\% in PBE. The problem of delocalization 
is thus not reduced in the many body framework and is rather substantial in the $GW$ approximation - 
similar to the RPA. The density matrix analysis for H$_2$ (Sec.~\ref{sec-2level}) in a minimal basis shows that we correctly obtain 
half occupation of the natural orbitals in the dissociation limit. This suggest that also in LiH we may 
dissociate correctly in $GW$. However, again this would occur at an unphysically large $R$.

\section{Summary and Conclusions}\label{sec:conclusion}
In summary, we have presented a general assessment of the sc$GW$ 
and scRPA approximations for ground-state properties of prototype diatomic 
molecules composed of open shell atoms.
The analysis of the different total energy contributions reveals 
large differences between the use of a non-local and frequency 
dependent self-energy in the $GW$ approximation and the use of 
a static and local potential in RPA, both evaluated at self-consistency. 

Non-locality leads to smaller Coulomb correlation energies, due 
to the larger gap between occupied and empty states. 
From the analysis of the density matrix, we deduce that 
the sc$GW$ ground-state is not representable 
in terms of a single Slater determinant.
The non-integer eigenvalues of the density matrix illustrates a transition from 
a non-interacting to an interacting ground state at self-consistency. 
Furthermore, screening effects within $GW$ increase the kinetic energy 
as compared to Hartree-Fock. A similar effect can be achieved using the static local
RPA potential.

The solution of a two-level model of the H$_2$ molecule at dissociation 
complements our first-principles calculations and allows to investigate the 
limit of infinite interatomic separation. For this simplified model, 
both $GW@{\rm HF}$ and RPA reach the correct dissociation limit. 
However, within $GW$ the bond-breaking regime is reached at an unphysical 
large bond length. 

For the case of LiH, we have shown that both scRPA and sc$GW$ are affected by 
strong delocalization or fractional charge errors. In RPA this leads to the dissociation of 
LiH in non-neutral fragments. 

The sc$GW$ dissociation curve can be considerably improved 
by resorting to spin-unrestricted calculations, whereas 
in scRPA, spin-unrestricted and restricted coincide at dissociation.
This demonstrates that scRPA --  similarly to perturbative RPA -- is free 
from fractional-spin error. On the other hand, sc$GW$ suffers from 
a large static correlation or fractional spin errors that is responsible for the qualitatively 
incorrect description of the dissociation region.

\acknowledgements
AR acknowledges financial support by the European Research Council Advanced Grant DYNamo (ERC-2010- AdG-267374), 
European Commission project CRONOS (Grant number 280879-2 CRONOS CP-FP7), Spanish Grant (FIS2013-46159-C3-1-P) and 
Grupo Consolidado UPV/EHU del Gobierno Vasco (IT578-13).
PR acknowledges financial support from the Academy of Finland through its Centres of Excellence Program (project no. 251748).
%
\end{document}